# Presolar Stardust in Highly Pristine CM Chondrites Asuka 12169 and Asuka 12236

Larry R. Nittler[1], Conel M. O'D. Alexander[1], Andrea Patzer[1,2], and Maximilien J. Verdier-Paoletti[1,3]

[1]Earth and Planets Laboratory, Carnegie Institution of Washington, 5241 Broad Branch Rd NW, Washington, DC 20015, USA.

[2]Geosciences Center Göttingen, University of Göttingen, Goldschmidtstr. 1, 37077 Göttingen, Germany

[3]Institut de Minéralogie, de Physique des Matériaux, et de Cosmochimie (IMPMC), Sorbonne Université, Muséum national d'Histoire naturelle, UPMC Université Paris 06, UMR CNRS 7590, IRD UMR 206, 75005 Paris, France

**Abstract**

We report a NanoSIMS search for presolar grains in the CM chondrites Asuka (A) 12169 and A12236. We found 90 presolar O-rich grains and 25 SiC grains in A12169, giving matrix-normalized abundances of 275 (+55/-50, 1σ) ppm or, excluding an unusually large grain, 236 (+37/-34) ppm for O-rich grains and 62 (+15/-12) ppm for SiC grains. For A12236, 18 presolar silicates and 6 SiCs indicate abundances of 58 (+18/-12) and 20 (+12/-8) ppm, respectively. The SiC abundances are in the typical range of primitive chondrites. The abundance of presolar O-rich grains in A12169 is essentially identical to that in CO3.0 Dominion Range 08006, higher than in any other chondrites, while in A12236 it is higher than found in other CMs. These abundances provide further strong support that A12169 and A12236 are the least-altered CMs as indicated by petrographic investigations. The similar abundances, isotopic distributions, silicate/oxide ratio, and grain sizes of the presolar O-rich grains found here to those of presolar grains in highly primitive CO, CR and ungrouped carbonaceous chondrites (CCs) indicate that the CM parent

body(ies) accreted a similar population of presolar oxides and silicates in their matrices to those accreted by the parent bodies of the other CC groups. The lower abundances and larger grain sizes seen in some other CMs are thus most likely a result of parent-body alteration and not heterogeneity in nebular precursors. Presolar silicates are unlikely to be present in high abundances in returned samples from asteroids Ryugu and Bennu since remote-sensing data indicate that they have experienced substantial aqueous alteration.

**Introduction**

Carbonaceous chondrite meteorites (CCs) accreted in the Sun's protoplanetary disk and provide a record of the earliest planet-formation processes. It is desirable to identify meteorites with the least amount of parent-body modification (e.g., aqueous alteration, thermal metamorphism) because these may provide a largely unmodified sample of presolar and nebular materials. The most abundant CCs, Mighei-type or CM, show a wide range of parent-body modification, but all have been altered to some extent. Moreover, with the impending return of samples from asteroids Ryugu and Bennu by the Hayabusa2 and OSIRIS-REx spacecraft, respectively, CM chondrites have come under increasing scrutiny. For example, spectroscopic evidence suggests both bodies may be related to this class of meteorites (Hamilton et al., 2019; Kitazato et al., 2019).

Until quite recently, the Paris CM meteorite was considered to be the least altered and hence most primitive known CM chondrite. Hewins et al. (2014) estimated its petrologic subtype to be 2.9, based on the scale of Rubin et al. (2007), whereas Marrocchi et al. (2014) estimated it to be 2.7. In any case, the degree of alteration in Paris is heterogeneous, with some regions clearly exhibiting, for example, more FeNi metal than others (Verdier-Paoletti et al., 2020). Oxidation of FeNi metal occurs very early in the process of aqueous alteration. Compared to most CMs, Paris, especially the least-altered portions, is characterized by lower abundances of phyllosilicates and other obvious products of aqueous alteration and its matrix contains amorphous silicate materials (Leroux et al., 2015) bearing similarity to GEMS (Glass with Embedded Metal and Sulfides) grains that are found in highly primitive interplanetary dust particles (Bradley, 1994; Keller and Messenger, 2011). Recent studies have shown that three Antarctic CM chondrites are even more pristine than Paris (Kimura et al., 2019; Kimura et al., 2020; Noguchi et al., 2020; Tsuchiyama et al., 2020). These meteorites – Asuka 12085, Asuka 12236, and Asuka 112169 (hereafter, A12085, etc.) – contain abundant FeNi metal and show very little evidence of aqueous alteration (e.g., phyllosilicates or tochilinite-cronstedtite intergrowths) or heating (based on Raman spectroscopy). Kimura et al. (2020) estimate their petrologic subtypes as 2.8 (A12085), 2.9 (A12236), and 3.0 (A12169). More recently, Glavin et al. (2020) found a much higher abundance of amino acids in A12236 than in Paris, providing further support that this meteorite has been less affected by parent-body alteration than any previously studied CMs.

One sensitive measure of the alteration state of primitive chondrites is the abundance of presolar stardust grains. Presolar grains are rare (10s – 100s ppm), generally sub-µm dust grains with extremely unusual isotopic compositions in most or all elements they contain. Their isotopic compositions indicate an origin in evolved stars (e.g., red giants, supernovae) prior to the formation of the Sun. They were part of the protosolar molecular cloud and, as pristine stardust, provide useful information on a whole array of astrophysical and cosmochemical processes (e.g., see reviews by Zinner, 2014; Nittler and Ciesla, 2016). A wide range of presolar phases, including oxides, silicates, carbide, nitrides, and elemental C, have been identified. Because these respond in different ways to heating and aqueous alteration processes in asteroids, their relative abundances can be used for evaluating the degree of metamorphism/alteration and tracking such processes in meteorite parent bodies (Huss and Lewis, 1995; Huss et al., 2003; Leitner et al., 2012; Davidson et al., 2014). For example, presolar SiC and $Al_2O_3$ are destroyed by thermal metamorphism and are thus absent or in very low abundance in heated chondrites, but are less affected by aqueous alteration and present in even highly altered meteorites like CI1 Orgueil (Hutcheon et al., 1994; Huss and Lewis, 1995). Presolar silicate grains are particularly susceptible to destruction by asteroidal processes and thus are generally only found in the least altered extraterrestrial samples. The highest abundances of presolar silicates have been reported in some cometary interplanetary dust particles (>400 ppm; Floss et al., 2006; Busemann et al., 2009). Slightly lower abundances (100-240 ppm) have been reported for Antarctic micrometeorites, and the matrices of the ungrouped C3 Acfer 094 and the least altered CR and CO chondrites (Floss and Haenecour, 2016a), with the highest abundance reported to date of 240 ppm for the CO3.0 Dominion Range (DOM) 08006 (Haenecour et al., 2018; Nittler et al., 2018). Note that all quoted abundances are matrix-normalized.

CM chondrites have a similar abundance (10s of ppm) of presolar SiC in their matrices to other primitive chondrites and, in fact, the vast majority of published data on SiC comes from the CM2 Murchison meteorite (Amari et al., 1994). In contrast, CM chondrites have been found to have lower abundances of presolar silicates than CRs, CO3s, and Acfer 094. Zhao et al. (2014) identified no presolar silicates in the CM-related Sutter's Mill meteorite. Leitner et al. (2020) recently reported the most in-depth search for presolar silicates in CMs to date. They searched fine-grained chondrule rims of six CMs and Sutter's Mill, and found a combined abundance of O-rich presolar grains (silicates and oxides) of 18 ppm. The Leitner et al. (2020) study did not include the less-altered Paris CM chondrite. An initial search of a small area of this meteorite found only an upper limit of 10 ppm for the abundance of presolar silicates (Mostefaoui, 2011). More recently, Verdier-Paoletti et al. (Verdier-Paoletti et al., 2019, 2020) performed a more extensive NanoSIMS survey of Paris. Verdier-Paoletti et al. (2020) reported an average abundance of presolar O-rich grains of ~37 ppm, with a slightly higher abundance of ~50 ppm in areas that contain more FeNi metal and thus appear to be less altered. Nittler et al. (2019) reported a similar abundance of 31 ppm for O-rich stardust in Northwest Africa 5958, a primitive ungrouped CC that may be related to CMs (Jacquet et al., 2016).

We report here a NanoSIMS-based search for presolar grains in the two least altered CM chondrites reported by Kimura et al. (2020): A12236 and A12169. High abundances of presolar SiC and silicate grains in both meteorites confirm that they are indeed highly pristine; A12169 in particular has at least as high an abundance of presolar O-rich grains as CO3.0 DOM 08006, heretofore the meteorite with the highest measured abundance. Additional NanoSIMS characterization of the organic matter in both meteorites (Nittler et al., 2020a; 2020b) will be reported in detail elsewhere.

**Samples and Methods**

We obtained polished thin sections (PTS) of A12236 (Fig. 1) and A12169 (Fig. 2) from the Japanese National Institute of Polar Research. The A12169 PTS is the same one investigated by Kimura et al. (2020) and Noguchi et al. (2020). We analyzed both the A12236 and A12169 PTS by point-counting in a JEOL 8530F electron probe microanalyzer (EMPA, at 15 keV, 5 nA, 3 μm spot size, applying 150 µm grids) and by energy dispersive x-ray spectroscopy (EDS) with a JEOL 6500F scanning electron microscope (SEM). The results of the point-counting studies are only briefly discussed here to provide a general petrologic context and will be thoroughly addressed elsewhere.

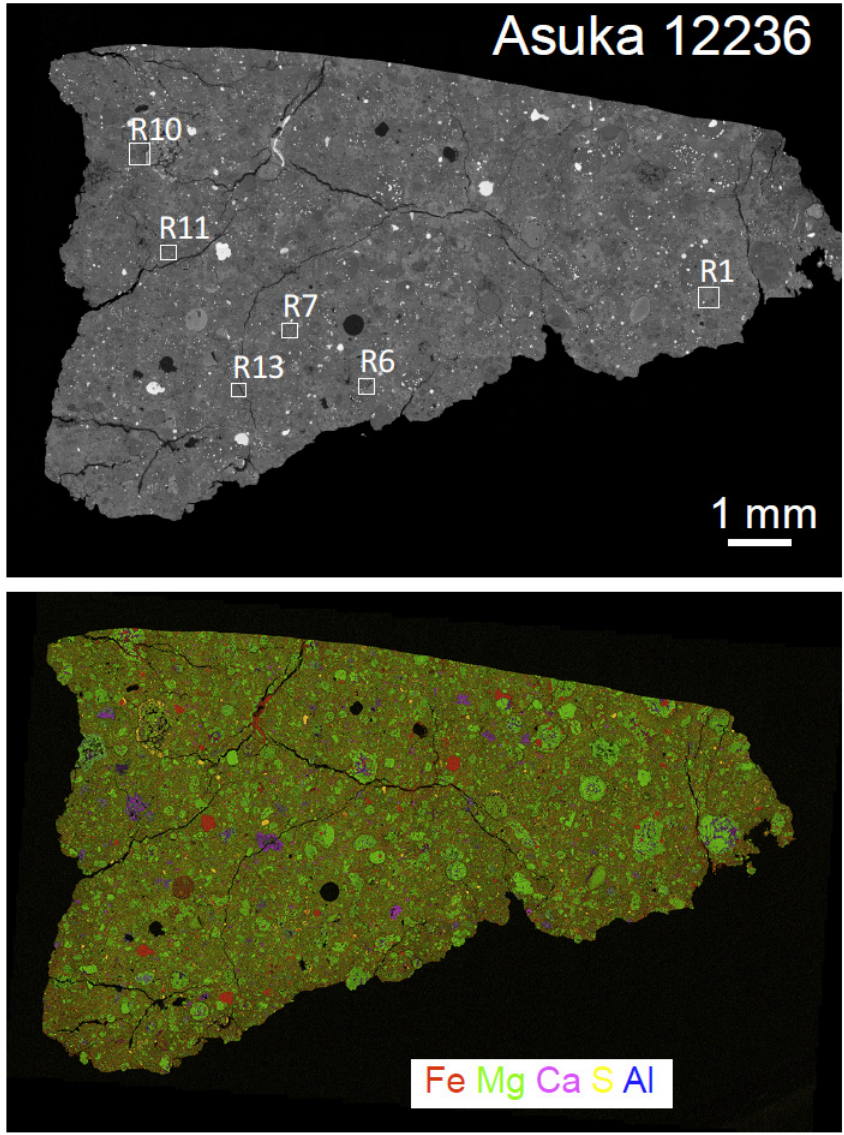

Fig. 1: Maps of Asuka 12236 polished thin section. Top: Backscattered electron map. Labeled boxes indicate regions of matrix targeted for NanoSIMS analysis (Fig. 3). Bottom: False-color combined EDS element map (Red=Fe, Green=Mg, Magenta=Ca, Yellow=S, Blue=Al).

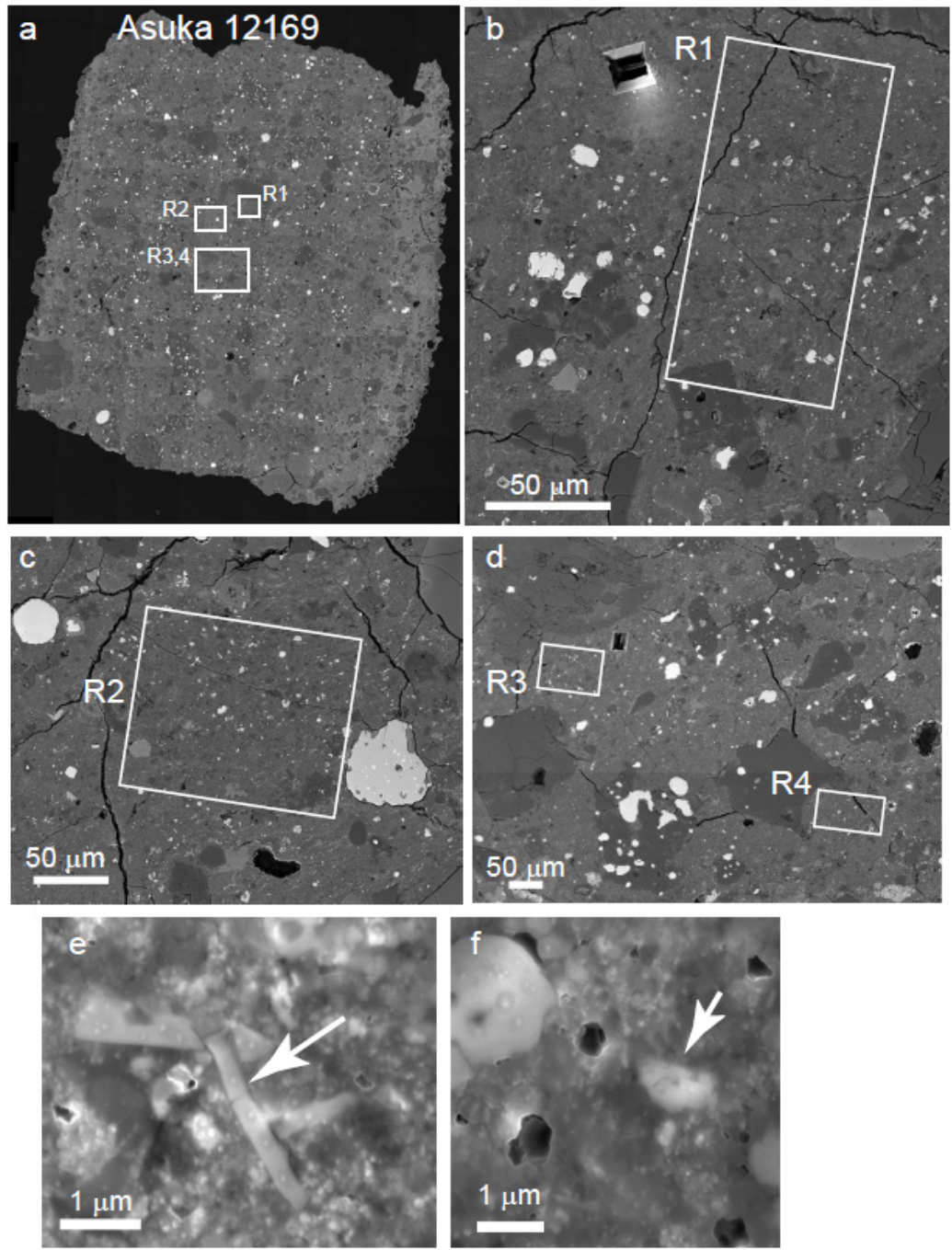

Fig. 2. a) Back-scattered electron map of Asuka 12169 polished thin section. The labeled boxes indicate regions selected for NanoSIMS analysis and shown at higher magnification in panels b-d. e) Secondary-electron image of a 2-μm by 250-nm enstatite needle in A12169 (arrow). f) Secondary-electron image of a large presolar hibonite grain, A12169-23 (arrow).

Based on the EMPA and SEM analyses, multiple areas of both Asuka meteorites were selected for NanoSIMS presolar grain searches (Figs. 1-3). We used standard methods for automatic isotopic imaging with a Cameca NanoSIMS 50L in multicollection mode (Nittler et al., 2018). Namely, a finely focused $Cs^+$ primary ion beam was rastered over 15×15 µm² (A12236) or 10×10 µm² (A12169) areas with synchronized collection of 256×256 pixel images of negative secondary ions of $^{12,13}C$, $^{16,17,18}O$, $^{28}Si$, and $^{27}Al^{16}O$ on electron multipliers, as well as secondary electrons. The charge-neutralization electron gun was not used since the relatively high C contents and fine-grained nature of the meteorite matrix generally kept the samples from charging. A primary beam intensity of ~1 pA was selected to keep the maximum $^{16}O^-$ beam intensity at ~500,000 counts per second. The entrance and exit slits of the mass spectrometer were set to allow resolution of $^{13}C$ from $^{12}CH$ and of $^{17}O$ from $^{16}OH$. Prior to each measurement, a 150-pA beam was rastered over a slightly larger area to remove the C coat and achieve stable secondary count rates (about 5 minutes). Each measurement consisted of 20 (A12236) or 18 (A12169) repeated cycles with counting times of 197 s and 98 s per cycle for A12236 and A12169, respectively. These conditions resulted in a total of ~17.5 s counting time per square micron for both meteorites. After each measurement, the stage was automatically moved to an adjacent location and the process repeated, with a few micrometers left in between images to preserve some C coat and thus reduce the possibility of sample charging.

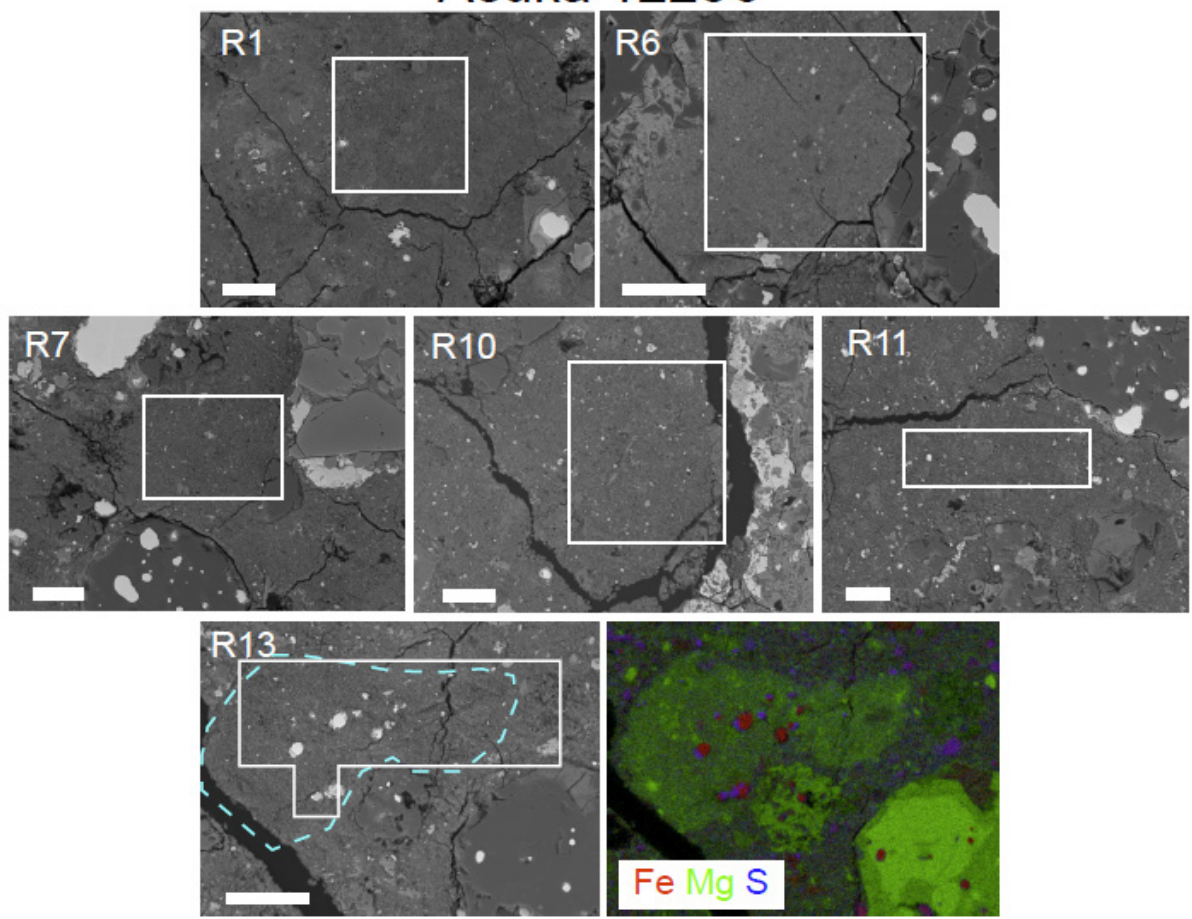

Fig. 3. Asuka 12236 matrix regions analyzed by NanoSIMS. All panels except bottom right are backscattered electrons. Bottom-right panel is an RGB composite image of EDS elemental maps for region R13 (Red=Fe, Green= Mg, Blue=S). The cyan outline in the bottom left panel indicates boundary of unusually Mg-rich matrix in region R13.

We used the IDL (L3Harris Geospatial Solutions, Inc.) based L'image software package (L. R. Nittler, Carnegie Institution) to analyze the images, following the methods described by Nittler et al. (2018). We corrected the images for cycle-to-cycle image shifts, the 44-ns deadtime of the NanoSIMS pulse-counting system (all isotopes), and quasi-simultaneous arrival (QSA, Slodzian et al., 2004) effects (just the $^{16}O$ images). Deadtime and QSA corrections were made pixel-by-pixel according to the instantaneous count-rate of each isotope. The proportionally constant relating measured isotope ratio to ratio of secondary to primary ion intensity needed for the QSA correction was determined empirically from $^{18}$O/$^{16}$O ratio images for different image runs. It typically ranged from 0.5 (the expectation for Poisson statistics) and 1.0. For comparison, Jones et al. (2017) recently measured this factor to be 0.7 when measuring O and S isotopes with a Cameca ims75/GEO ion probe.

Following image corrections, we computed pixel-by-pixel C and O ratio images. The ion images were smoothed with a 3×3 pixel boxcar prior to calculating ratio images to boost signal-to-noise ratios and eliminate high-frequency image artifacts. Oxygen isotopes were internally normalized to the average composition of each image and C isotopes were calibrated based on measurements of external standards (synthetic SiC and isolated insoluble organic matter, or IOM, from the Queen Alexandra Range [QUE] 99177 meteorite). Candidate presolar grain regions of interest (ROIs) were selected as regions of several contiguous pixels whose $^{13}$C/$^{12}$C, $^{17}$O/$^{16}$O, or $^{18}$O/$^{16}$O ratios clearly differed from the average values measured in a given image, as described by Nittler et al. (2018). Isotopic ratios were calculated from the total counts within ROI outlines. Errors were determined from counting statistics, e.g., based on the square root of the total number of counted secondary ions. For grains showing isotopic depletions (e.g., in $^{17}$O, $^{18}$O, or $^{13}$C), the counting statistical error was calculated from the total counts expected for a terrestrial isotopic ratio. We required a significance level of 4.5σ for anomalous isotope ratios in order for a grain to be considered presolar. Note that Hoppe et al. (2015) and Leitner et al. (Leitner et al., 2020) have used the approximation formulae of Gehrels (1986) to determine confidence levels for isotope measurements from NanoSIMS images and used a 5.3σ threshold to accept a grain as presolar, based on measurements of a terrestrial sample. All but three of the presolar grains reported in this work exceed this threshold when the Gehrels (1986) formulae are used to calculate significance levels. The three exceptions have moderate $^{18}$O depletions and based on their anomalies being present through all measurement cycles, we are confident that these are indeed presolar grains.

Following the NanoSIMS measurements, some of the analyzed meteorite areas were re-examined with the JEOL 6500F SEM to attempt to identify presolar grains and determine their composition by EDS, if possible. However, as discussed later, most of the grains are smaller than 300 nm in diameter and thus difficult to get EDS data for without strong contributions from the surrounding material. Thus, for most of the identified grains, we used the NanoSIMS images to only roughly estimate their mineralogies, e.g., silicate versus oxide.

## Results

SEM and EMPA analysis of the A12236 section gave very similar results to those reported by Kimura et al. (2020) for a different section of this meteorite. Namely, it has a typical CM texture and mineralogy (Fig. 1), with 59.0 vol.% matrix, 32.8 vol.% chondrules, 3.7% refractory inclusions (RIs), and the remainder isolated (>10-µm) silicate (2.7 vol.%) and metal/sulfide grains (1.7 vol.%). For comparison, Kimura et al. (2020) found 64 vol.% matrix, 28.9 vol.% chondrules, and 3.8 vol.% refractory inclusions. Rare calcite grains are present in the matrix, but no tochilinite-cronstedtite intergrowths (TCI) were observed, and a few (≈150-µm) CI-/CM-like clasts were also seen. We selected several different matrix regions for NanoSIMS analysis based on slight differences in texture and/or chemistry (e.g., different amounts of metal or sulfide grains). For example, region R13 contains a small area for which an SEM-EDS map (Fig. 2, bottom right) indicates clearly higher Mg contents than surrounding materials.

For the A12169 PTS, we observed very similar volume fractions of chondritic components, namely 56.5% matrix, 35.6% chondrules, 3.0% RIs, 2.0% isolated silicate grains and 2.4% isolated opaque grains. The higher abundance of isolated metal/sulfide grains when compared to A12236 may be a true feature of the more pristine A12169 sample. One must bear in mind, however, the diminished accuracy of our point-counting technique (150 µm grids) when it comes to components with small sizes and in low abundances. Within this uncertainty, we deem the fractions of RIs and isolated grains very similar, if not indistinguishable. However, because it is the same section studied by Kimura et al. (2020) and Noguchi et al. (2020), we briefly review some of their observations here. See Fig. 1c of Kimura et al. (2020) for a combined elemental (Mg-Ca-Al) map of this section. As in A12236, TCIs have not been seen in A12169, though some 2-3 µm patches of fibrous material are present and hypothesized to be poorly crystallized tochilinite, perhaps precursors of TCI (Noguchi et al., 2020). The matrix consists of sub-µm silicates (likely amorphous), Fe-Ni metal, and abundant <100-nm Fe sulfides. Although Noguchi et al. (2020) reported SEM evidence for phyllosilicates on the surfaces of some chondrules, they are not apparently present in the matrix. Based on our point-counting data, some analyses of chondrules are consistent with the presence of chrysotile (normative), in particular in the chondrules of A12236 (4.5% of chondrule analyses vs. 0.8% in A12169). Noguchi et al. (2020) and Tsuchiyama et al. (2020) reported that enstatite whiskers and platelets are common in the matrix of A12169 based on TEM and synchrotron-based nanotomography measurements and we observed some enstatite whiskers in the SEM as well (e.g., Fig. 2e). We selected regions near the center of the PTS (Fig. 2a) to avoid an obviously altered area near the edge of the section. The selected regions are all fine grained with higher metal/sulfide abundances than seen in most of the A12236 areas. Two regions are adjacent to locations where FIB sections were removed prior to our receiving the PTS.

The results of the presolar grain searches are summarized in Table 1. We mapped a total of ≈16,000 μm$^2$ in A12236 and ≈24,000 μm$^2$ in A12169, divided among the various regions as indicated. We identified a total of 18 presolar O-rich and 6 presolar SiC grains in A12236 and 90 presolar O-rich and 25 presolar SiC grains in A12169. SiC was identified on the basis of anomalous $^{13}$C/$^{12}$C ratios associated with Si. Isolated SiC grains usually show Si/C secondary ion ratios of close to unity, The SiC grains identified here have Si$^-$/C$^-$ ratios in the range of ~0.2-1.2; a few relatively low ratios are due to dilution by surrounding carbonaceous material. A few anomalous ($\delta^{13}$C ~ -300 to +300 ‰) C-rich ROIs were identified without associated Si in both meteorites. These are most likely organic in nature and will be discussed elsewhere, along with a more extensive survey of organic matter in these samples. Data for the O-rich and SiC grains are provided in Tables 2 and 3, respectively, and O isotope data for the O-anomalous grains are compared with literature data in Fig. 4. The mineralogies of the O-rich phases were determined by SEM-EDS for a few grains, but mostly were estimated from Si$^-$/O$^-$ and AlO$^-$/O$^-$ secondary ion ratios determined from the NanoSIMS images. Note that such assignments can be unreliable (Nguyen et al., 2007; 2010). However, based on the large body of extant data on presolar stardust in primitive CCs from Auger spectroscopy (Floss and Haenecour, 2016a), it is highly likely that a substantial fraction of the identified grains is made up of silicates rather than oxides. The $^{12}$C/$^{13}$C ratios of the 30 presolar SiC grains range from 16 to 146, with an average value of ≈54.

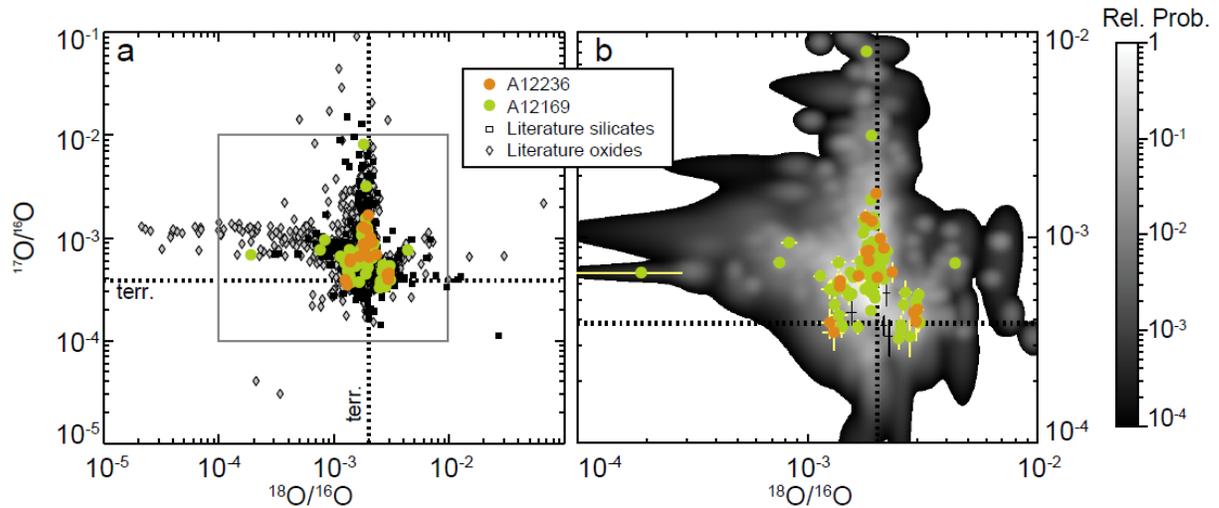

Fig. 4. a) The O isotopic compositions of presolar O-rich grains identified in A12236 and A12169 are compared to literature data for presolar oxides and silicates in meteorites. The box indicates the plot range of panel b. b) A zoomed-in plot of the Asuka grain compositions overlaid on a probability density map of the literature in situ data. See text for details. The dashed lines indicate terrestrial (≈bulk meteorite) isotope ratios. Literature data are taken from a large number of sources (see, e.g., Nittler et al., 2008; Nguyen et al., 2010; Leitner et al., 2012; Zinner, 2014; Floss and Haenecour, 2016).

The sizes (effective diameters) of the presolar grains were calculated from the ion images and corrected for 100-nm beam broadening as described in detail by Nittler et al. (2018). Histograms of the presolar O-rich grain sizes are shown in Fig. 5. The average sizes of the presolar O-rich grains are similar in both meteorites (~240 nm for A12236, 270 nm for A12169) and are very similar to those seen in many other carbonaceous chondrites (e.g., summarized by Leitner et al., 2020). The A12169 data set contains seven O-rich presolar grains larger than 400 nm in diameter, whereas no such relatively large grains are seen in A12236. The large grains make up ~8% of the A12169 grains, similar to the fraction of ~11% for >400-nm presolar O-rich grains in DOM 08006 and Acfer 094 (e.g., Mostefaoui and Hoppe, 2004; Nguyen et al., 2007; Vollmer et al., 2009; Hoppe et al., 2015; Nittler et al., 2018; Haenecour et al., 2018). Thus, one would expect to find 1–2 such grains among the A12236 population. Their lack thus may be simply due to the poor statistics. Alternatively, the lack of larger grains may indicate a reduction in size due to progressive alteration, though the observation by Leitner et al. (2020) that presolar grains in the more altered CM chondrites are on average larger than in less-altered chondrites argues against this. A larger data set in A12236 could help explain this difference. Few O-rich presolar grains as large as 1-μm have been found in previous in situ searches (e.g., Leitner et al., 2012; Leitner et al., 2017). The largest grain identified here, A12169-23 (Fig. 2f), is about 1.1 μm across; SEM-EDS revealed it to be a Cr-bearing hibonite ($CaAl_{12}O_{19}$). The average size of the identified SiC grains (Table 3) is 250 nm in both meteorites, essentially identical to that found by *in situ* searches of primitive CR, CO, and ungrouped CCs (e.g., summarized by Leitner et al., 2020).

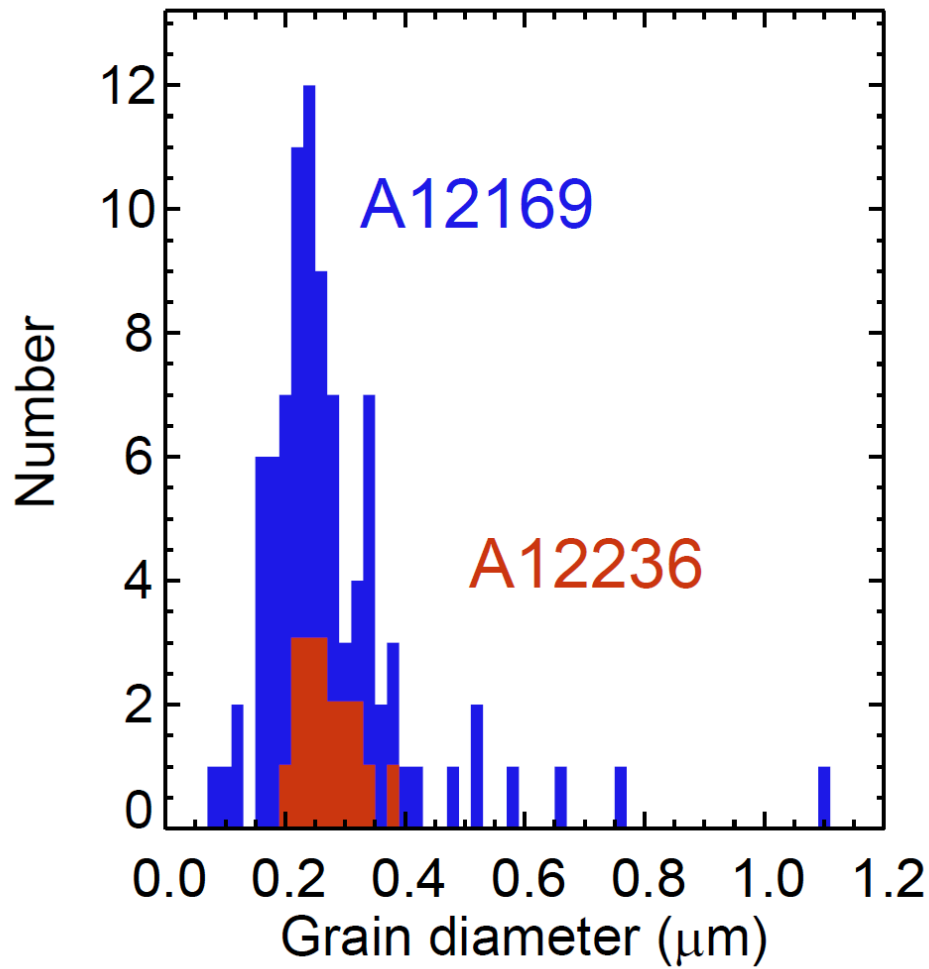

Fig. 5. Histograms of the effective diameters of identified presolar O-rich grains in A12169 and A12236.

We calculated the matrix-normalized abundances (Table 1) of the presolar grains in the two meteorites by dividing the total area occupied by the identified presolar grains by the total scanned area in each analyzed region. The abundance errors reported for individual regions are 1-sigma and are based on the number of identified grains and the confidence limits for small numbers of events tabulated by Gehrels et al. (1986). However, to estimate the errors on the overall abundances for each meteorite, we used the Monte Carlo method of Nittler et al. (2018). This method provides more realistic error estimates as it takes into account the variable sizes of the identified grains in addition to the uncertainty due to the number of grains. Note also that the presence of unusually large grains can skew abundance estimates. For example, a 1.1-µm-diameter grain like A12169-23 has the same cross-sectional area as >16 grains of typical, ~270-nm, size. We thus performed two abundance calculations for the A12169 dataset, one with all of the grains and one excluding A12169-23 (Table 1). As can be seen in Table 1 and Fig. 6, excluding the big grain lowers the overall abundance estimate for O-rich presolar grains from 275 ppm to 236 ppm.

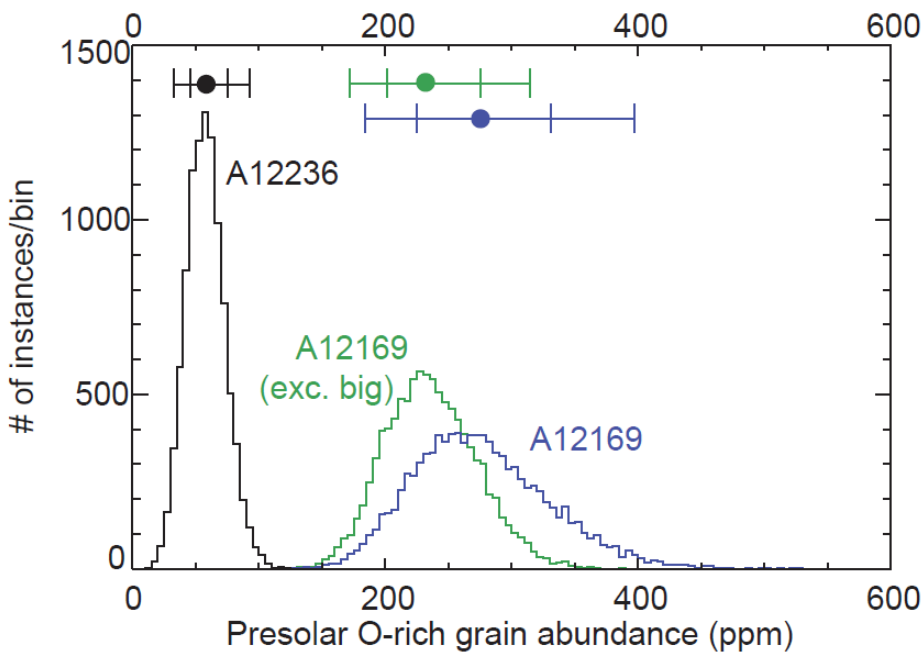

Fig. 6. Results of Monte Carlo modeling (Nittler et al., 2018) of presolar O-rich grain abundances in A12169 and A12236. Two results are provided for A12169, one based on the 90 identified presolar grains in this meteorite and one ("A12169 exc. big") that excludes the unusually large grain A12169-23. Histograms indicate the distributions of abundances for 10,000 MC instances for each calculation while circles and error bars indicate average abundances and ±1-σ and 2-σ error bars (84.1% and 97.7% enclosed probability).

**Discussion**

The O isotopes of the 108 presolar O-rich grains identified in A12236 and A12169 are compared to literature data in Fig. 4a. While the new grains overlap with the literature data, they span a narrower isotopic range, e.g., with few grains exhibiting large $^{18}O$ depletions ($^{18}O/^{16}O < 10^{-3}$). This difference has long been recognized as reflecting the contribution of signal from neighboring material to the presolar grains during *in situ* NanoSIMS imaging measurements due to the tails of the primary ion beam (Nguyen et al., 2007). This "isotopic dilution" leads to measured ratios being less anomalous than their true intrinsic values. In contrast, the literature data includes a large number of oxide grains from meteoritic acid residues (e.g., Nittler et al., 2008) that were measured as isolated grains and were thus less susceptible to contamination. In principle, one can correct for this dilution via modeling but we have not done so here as our conclusions do not depend on the precise isotopic compositions of any of the grains. To compare the current grains with literature data that are also subject to isotope dilution, we overlay them in Fig. 4b on a probability density map determined for just the literature data acquired by *in situ* NanoSIMS imaging of meteorite thin sections. This map is essentially a 2-dimensional histogram including error weighting and was calculated by assigning each grain a 2-dimensional Gaussian probability distribution based on the reported uncertainties in its isotopic ratios and summing all of the individual distributions. The new presolar grains span very similar ranges to the literature in situ data; their classification into the Group definitions of Nittler et al. (1997) is given in Table 2. The peak of the literature probability distribution corresponds to the Group 1 grains, with $^{17}O$ enrichments and solar-to slightly depleted $^{18}O/^{16}O$ ratios, and a large fraction of the new grains overlap with this peak, though a cluster of moderately $^{18}O$-enriched Group 4 grains and a smattering of grains with more extreme anomalies are seen as well. The interpretation of the O-isotopic data in terms of likely stellar sources of the grains and implications for stellar evolution, nucleosynthesis and Galactic chemical evolution has been discussed at length in the literature (see, e.g., Nittler et al., 2008; Zinner, 2014) and we do not repeat those discussions here. For our purposes, it is sufficient to note that the Asuka CM meteorites contain a very similar isotopic mix to that seen in other primitive chondrites.

As with the O-rich presolar grains, the presolar SiC grains identified in A12236 and A12169 show a very similar range of $^{12}C/^{13}C$ ratios to literature data. Presolar SiC has been classified into numerous sub-groups based on their multi-element isotopic compositions (e.g., Zinner, 2014). With only the C isotopic ratios, we cannot uniquely classify the Asuka grains. However, the vast majority of them, with $^{12}C/^{13}C$ ratios between 20 and 100 most likely belong to the dominant "mainstream" population, the two grains with $^{12}C/^{13}C$ >120 could be either X or Y grains, and the two grains with $^{12}C/^{13}C$ <20 could be either mainstream grains or AB grains whose true C ratios of lower than 10 have been affected by isotopic dilution during the NanoSIMS measurement. As for O-rich grains, the reader is referred to the literature for details on the stellar origins and astrophysical implications of presolar SiC grains (see, e.g., Hoppe et al., 2000; Zinner et al., 2006; Zinner, 2014; Liu et al., 2018).

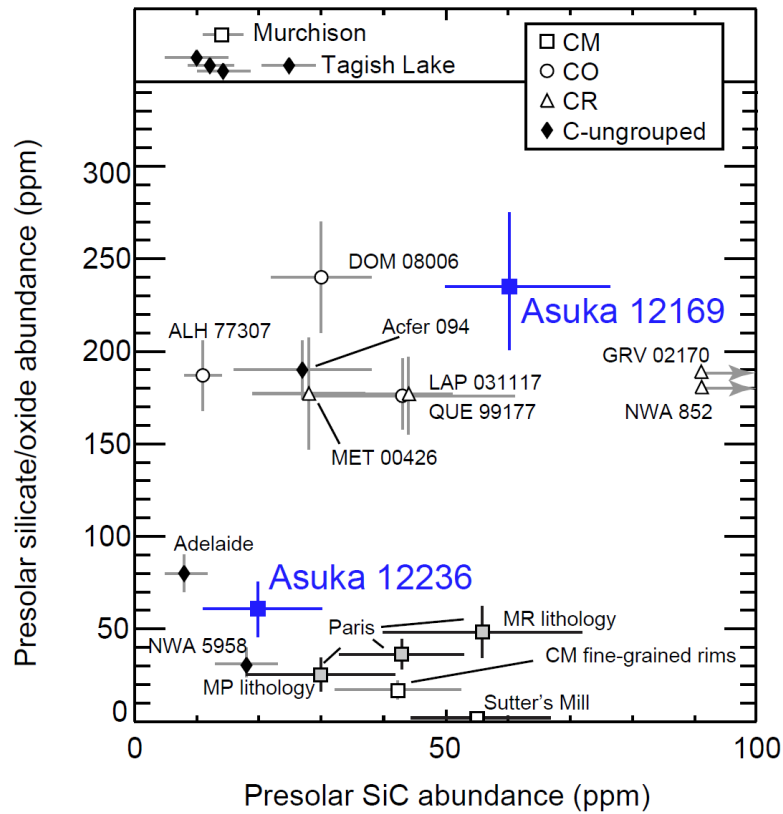

Fig. 7. Abundances of presolar grains in A12236 and A12169 are compared those in other chondrites. ALH= Allan Hills. GRA=Grave Nunataks. GRV=Grove Mountain. LAP=La Paz Icefield. MP=metal-rich. MP=metal-poor. NWA=Northwest Africa. QUE=Queen Alexandra Range. Data sources: (Floss and Stadermann, 2009; Nguyen et al., 2010; Floss and Stadermann,

2012; Leitner et al., 2012; Zhao et al., 2013; Davidson et al., 2014; Zhao et al., 2014; Haenecour et al., 2018; Nittler et al., 2018; Nittler et al., 2019; Riebe et al., 2019; Leitner et al., 2020; Verdier-Paoletti et al., 2020)

The matrix-normalized abundances of presolar O-rich grains are plotted against those of presolar SiC for A12236, A12169, and a number of other primitive CCs in Fig. 7. All abundances of presolar O-rich grains are taken from *in situ* NanoSIMS imaging searches carried out under similar conditions to the present study, with spatial resolution of order 100-150 nm. Note that Hoppe et al. (2015, 2017) have used higher-resolution measurements to show that the primitive chondrites Acfer 094, QUE 99177, and MET 00426 all contain a significant amount of O-rich presolar grains that are missed under standard conditions. Therefore, all plotted abundances of O-rich grains should be considered lower limits, but we can still make meaningful comparisons of the different plotted meteorite results since all have been acquired under similar analytical conditions.

Both A12236 and A12169 have abundances of presolar SiC grains within the range seen in most other studied CCs. In terms of CMs, the SiC abundance in A12236 of 20 ppm overlaps with those of Murchison, the metal-poor (MP) lithology of Paris (Verdier-Paoletti et al., 2020), and the CM-related NWA 5958, whereas the higher abundance seen in A12169 (62 ppm) is similar to that of the Paris metal-rich (MR) lithology and Sutter's Mill. However, the SiC abundances for all the meteorites are estimated from limited numbers of grains and error bars are thus large and the significance of the observed differences is not very high. Only two meteorites have been reported to have SiC abundances substantially out of the range of those plotted in Fig. 7, Northwest Africa 852 (160 ppm; Leitner et al., 2012) and Grove Mountain 021710 (192 ppm; Zhao et al., 2013), both CR2s. The errors are large on both estimates and it is not yet known if these high abundances represent real differences to the vast majority of other chondrites or are statistical flukes.

Presolar O-rich grains show a much larger spread in abundance among primitive meteorites than do SiC grains. As discussed in Section 1, the highest abundances are seen in cometary interplanetary dust particles (>375 ppm; Floss et al., 2006; Busemann et al., 2009; Davidson et al., 2012). In meteorites, the highest abundances of 150-250 ppm are seen in the least altered CR, CM, CO, and ungrouped CCs, while all previously measured CMs, including the lightly altered Paris, have lower abundances. We note that high abundances have also been reported for two unequilibrated ordinary chondrites, Meteorite Hills 00526 and Queen Alexandra Range 97008, in conference abstracts (Floss and Haenecour, 2016b; 2016c), but the details have not yet been published. The O-rich presolar grain abundance in A12236, 58 ppm, is very similar to that reported for the least-altered, metal-rich lithology of Paris (Verdier-Paoletti et al., 2020), but higher than bulk Paris and other measured CMs. We note that the highest observed abundance for this meteorite, 93 ppm, was found for the Mg-rich matrix of region R13 (Fig. 3), but the statistical significance of this result is very low (Table 1). Strikingly, A12169 has a much higher abundance of presolar silicates and oxides (240 ppm) than A12236 that is essentially identical to that of DOM 08006, heretofore the meteorite with the highest measured average abundance. Note that this is a conservative estimate on the abundance since it was calculated by excluding the largest grain in the dataset. Interestingly, with either calculation of the A12169 abundance, the estimated 2-$\sigma$ lower limit on the abundance is ~160 ppm (Fig. 6), comparable to the typical abundances seen in ungrouped carbonaceous chondrite Acfer 094 and primitive CR chondrites. The high abundances of presolar O-rich grains in A12236 and, especially, A12169 strongly support the petrographic (Kimura et al., 2020; Noguchi et al., 2020) and organic chemical (Glavin et al., 2020) evidence that these samples have seen even less alteration than the Paris meteorite and that the matrix of A12169 is among the most primitive chondritic material available for study.

We note that following submission of this paper, a parallel study of presolar grains in A12169 was reported in an abstract by Xu et al. (2020). These authors obtained very similar results to ours, with a presolar O-rich grain abundance in this meteorite of 208±19 ppm and a SiC abundance of 73±12 ppm.

Floss and Stadermann (2009) first suggested that, in addition to overall abundances, the ratio of presolar silicates to oxides may be a useful tracer of parent-body aqueous alteration since phases like $Al_2O_3$ are more resistant to alteration than silicates, especially since many presolar silicates are amorphous and hence especially susceptible to destruction. Leitner et al. (Leitner et al., 2012; Leitner et al., 2020) followed up on this suggestion and calculated that the lower limit on the presolar silicate/oxide ratio in the protosolar cloud should be ~23, based on the assumed stellar sources of the grains and models of the relative production of different species in these sources. No meteorite has been identified with a ratio as high as this, consistent with the idea that some destruction of presolar silicates has taken place within the precursor materials of chondrite matrices, based on their lower abundances of presolar silicates compared to cometary dust. The small number of oxides identified in most *in situ* studies results in very large uncertainties in the silicate/oxide ratios. With the caveat that the determination of presolar grain mineralogy from NanoSIMS images is highly uncertain (Nguyen et al., 2007; 2010), we estimate that 5 of the 90 presolar O-rich grains in A12169 and zero of the 18 grains in A12236 are oxides. These numbers give 2-$\sigma$ lower limits on the silicate/oxide ratio of 9.1 and 7.7 for A12169 and A12236, respectively. These lower limits are similar to the reported ratios for the very primitive Acfer 094 and DOM 08006 meteorites and much higher than the ratio of 1.5 found by Leitner et al. (2020) for their sample of fine-grained rims in CM2 chondrites.

In addition to finding a lower overall abundance of presolar O-rich grains and a lower silicate/oxide ratio, Leitner et al. also reported a slightly larger average grain size for their presolar grains in CM2 chondrule rims than typically seen in primitive chondrites, ~330 nm versus ~270 nm. Given that even Paris shows a lower abundance of O-rich presolar grains compared to primitive CRs, COs and ungrouped CCs, they suggested that perhaps presolar silicates and oxides were heterogeneously distributed within the carbonaceous chondrite forming regions of the Sun's protoplanetary disk, with the precursor materials to CM chondrite matrices containing a distinctly lower abundance of presolar O-rich grains and a lower silicate to oxide ratio than those that formed the matrices of the other CC groups. In contrast, our results demonstrate that the matrix of the most primitive known CM chondrite, A12169, contains a very similar complement of presolar grains, in terms of both abundances and sizes, to other CC groups. This indicates that the lower abundances, silicate/oxide ratios, and larger grain size found in the more altered CM2s by Leitner et al. (2020) reflect more extensive processing by aqueous alteration on the meteorites' parent body(ies), not heterogeneity in the solar nebula.

The results on the Asuka meteorites suggest a rather rapid destruction of presolar silicates during the early stages of aqueous alteration in CM chondrites. The TEM observations of Noguchi et al. (2020) indicate that A12169 matrix is similar to the least altered matrix of Paris (Leroux et al., 2015), yet the former has some five times more presolar silicates than the latter. The significant difference in presolar grain abundances coincides with our results from the point-counting study (e.g., bulk densities, Al/Si/CI ratios of chondrules and matrix) that confirm the highest degree of primitiveness for A12169 followed by A12236, the less altered regions of Paris and the more altered regions of Paris. No TEM observations have been reported for A12236, but at the SEM scale it shows very few signs of aqueous alteration in the matrix (e.g., no phyllosilicates or TCI, only rare calcite grains) and X-ray diffraction analysis also indicated phyllosilicates are absent (Kimura et al., 2020). Kimura et al. (2020) estimate its petrologic sub-type at 2.8. It is thus somewhat surprising that our measurements show that A12236 has substantially lower abundances of presolar silicates than the lightly-altered CR chondrites QUE 99177 and Meteorite Hills 00426, since both of these meteorites contain phyllosilicates (Le Guillou and Brearley, 2014; Howard et al., 2015) and are classified as petrographic types <2.8 (Alexander et al., 2013; Howard et al., 2015). The A12236 section studied here shows some chemical and textural variability including the presence of CI/CM-like clasts and unusually Mg-rich matrix as seen in region R13 (Fig. 3; the bulk Mg/Si/CI matrix ratio of A12236, however, is slightly lower than that of A12169 suggesting heterogeneity in the distribution of Mg-bearing matrix phases). Our results do not show strong differences in presolar grain abundances between the different regions, but this comparison is hampered by poor statistics. TEM analysis of this meteorite and Paris, especially of the same areas targeted for NanoSIMS presolar grain searches, could be extremely useful for elucidating the processes by which low levels of alteration destroy presolar silicates as well as understanding the differences in the earliest stages of aqueous alteration between CM and CR chondrites.

## Summary and Conclusions

We have conducted a NanoSIMS-based search for presolar C- and O-rich grains in polished thin sections of the Asuka 12236 and Asuka 12169 CM carbonaceous chondrites. We identified presolar SiC, silicates, and oxides, with isotopic distributions within the range of those seen in similar surveys of other primitive meteorites. Both meteorites contain presolar SiC at similar matrix-normalized abundances (few 10s of ppm) to those seen in most other primitive meteorites, but higher abundances of presolar silicates and oxides than seen in any previously studied CM chondrite, including the lightly-altered Paris. The abundance of O-rich presolar grains in A12169, 240 ppm, is identical to that seen in the CO3.0 chondrite Dominion Range 08006. The lower limits on the presolar silicate/oxide ratio (8–9) in both meteorites are comparable to the ratios reported for other presolar-grain rich primitive meteorites and much higher than the value of 1.5 reported for CM2 fine-grained chondrule rims (Leitner et al., 2020). The average grain size for the presolar grains in both meteorites is also very similar to that seen for other primitive chondrites, and higher than reported for the CM2 chondrule rims. Our primary conclusions arising from these results are:

1) The close similarity in O- and C-isotopic compositions of the grains to those seen in other chondrites indicates that the region of the solar nebula where the CM chondrites formed sampled the same or a closely similar mixture of presolar grains to the formation regions of other CC groups.

2) The similar high abundance, grain size, and silicate/oxide ratio of O-rich presolar grains in A12169 to those seen in CO3.0 DOM 08006 indicates that the CM parent body(ies) accreted a similar population of presolar oxides and silicates in their matrices to those accreted by the parent bodies of the other CC groups. This indicates that the observed differences in these properties of O-rich presolar grains in fine-grained chondrule rims in CM2 chondrites (Leitner et al., 2020) are due to parent body alteration effects, not to heterogeneity in the chondrite forming regions of the solar nebula.

3) The higher abundances of presolar grains in A12236 and especially A12169 than seen in other CM chondrites provides further strong support to petrographic and chemical evidence (Glavin et al., 2020; Kimura et al., 2020; Noguchi et al., 2020) that these are the most pristine CM meteorites identified so far, and that A12169 is probably best classified as having a petrographic sub-type very close to 3.0.

4) The drop in O-rich presolar grain abundance from >200 ppm in A12169 to ~50 ppm in A12236 and the least-altered portions of the Paris meteorite indicates that presolar silicates are relatively rapidly destroyed during the earliest stages of aqueous alteration in asteroidal parent bodies of CM chondrites. Additional TEM and x-ray diffraction studies of primitive CMs are warranted to better understand the processes that destroy presolar grains and how these compare with those in CR chondrites and other CC groups.

5) Spacecraft-based observations of Ryugu and Bennu indicate that both C-type asteroids contain hydrated materials and may have affinities to CM chondrites (heated ones in the case of Ryugu). This suggests that the samples to be returned from these bodies by Hayabusa2 and OSIRIS-REx are unlikely to have high abundances of presolar silicates, since the presence of observable amounts of hydrated minerals implies a degree of aqueous alteration sufficient to destroy a large proportion of them. However, the spatial scale of the spacecraft observations is far larger than that of hand samples and the results so far thus do not preclude the presence of less altered material mixed in at finer scales. Systematic searches for presolar grains should certainly be carried out in samples from both asteroids both for the scientific power of the grains themselves and for their use as probes of parent-body processing.

**Acknowledgments:** We thank the NIPR for providing the meteorite samples and Jan Leitner and an anonymous referee for helpful comments on the manuscript. This work was supported by NASA grants NNX17AE28G to LRN and NNX80NSSC18K0599 to CMOD.

**References**

Alexander C. M. O'D., Howard K. T., Bowden R., and Fogel M. L. 2013. The classification of CM and CR chondrites using bulk H, C and N abundances and isotopic compositions. *Geochimica et Cosmochimica Acta* 123:244–260.

Amari S., Lewis R. S., and Anders E. 1994. Interstellar grains in meteorites. I. Isolation of SiC, graphite, and diamond: Size distributions of SiC and graphite. *Geochimica et Cosmochimica Acta* 58:459–470.

Bradley J. P. 1994. Chemically anomalous preaccretionally irradiated grains in interplanetary dust from comets. *Science* 265:925–929.

Busemann H., Nguyen A. N., Cody G. D., Hoppe P., Kilcoyne A. L. D., Stroud R. M., Zega T. J., and Nittler L. R. 2009. Ultra-primitive interplanetary dust particles from the comet 26P/Grigg-Skjellerup dust stream collection. *Earth and Planetary Science Letters* 288:44–57.

Davidson J., Busemann H., and Franchi I. A. 2012. A NanoSIMS and Raman spectroscopic comparison of interplanetary dust particles from comet Grigg-Skjellerup and non-Grigg Skjellerup collections. *Meteoritics & Planetary Science* 47:1748–1771.

Davidson J., Busemann H., Nittler L. R., Alexander C. M. O'D., Orthous-Daunay F.-R., Franchi I. A., and Hoppe P. 2014. Abundances of presolar silicon carbide grains in primitive meteorites determined by NanoSIMS. *Geochimica et Cosmochimica Acta* 139:248–266.

Floss C. and Haenecour P. 2016a. Presolar silicate grains: Abundances, isotopic and elemental compositions, and the effects of secondary processing. *Geochemical Journal* 50:3-25.

Floss C. and Haenecour P. 2016b. Meteorite Hills (MET) 00526: An unequilibrated ordinary chondrite with high presolar grain abundances. *Lunar and Planetary Science Conference* 47:Abstract #2276.

Floss C. and Haenecour P. 2016c. Presolar silicate abundances in the unequilibrated ordinary chondrites Meteorite Hills 00526 and Queen Alexandra Range 97008. *79$^{th}$ Annual Meeting of the Meteoritical society*, Abstract #6015.

Floss C. and Stadermann F. 2009. Auger Nanoprobe analysis of presolar ferromagnesian silicate grains from primitive CR chondrites QUE 99177 and MET 00426. *Geochimica et Cosmochimica Acta* 73:2415–2440.

Floss C. and Stadermann F. J. 2012. Presolar silicate and oxide abundances and compositions in the ungrouped carbonaceous chondrite Adelaide and the K chondrite Kakangari: The effects of secondary processing. *Meteoritics & Planetary Science* 47:992–1009.

Floss C., Stadermann F. J., Bradley J. P., Dai Z. R., Bajt S., Graham G., and Lea A. S. 2006. Identification of isotopically primitive interplanetary dust particles: A NanoSIMS isotopic imaging study. *Geochimica et Cosmochimica Acta* 70:2371.

Gehrels N. 1986. Confidence limits for small numbers of events in astrophysical data. *The Astrophysical Journal* 303:336–346.

Glavin D. P., McLain H. L., Dworkin J. P., Parker E. T., Elsila J. E., Aponte J. C., Simkus D. N., Pozarycki C. I., Graham H. V., Nittler L. R., and Alexander C. M. O'D. 2020. Abundant extraterrestrial amino acids in the primitive CM carbonaceous chondrite Asuka 12236. *Meteoritics and Planetary Science* 55:1979–2006.

Haenecour P., Floss C., Zega T. J., Croat T. K., Wang A., Jolliff B. L., and Carpenter P. 2018. Presolar silicates in the matrix and fine-grained rims around chondrules in primitive CO3.0 chondrites: Evidence for pre-accretionary aqueous alteration of the rims in the solar nebula. *Geochimica et Cosmochimica Acta* 221:379-405.

Hamilton V. E. Simon A. A. Christensen P. R. Reuter D. C. Clark B. E. Barucci M. A. Bowles N. E. Boynton W. V. Brucato J. R. Cloutis E. A. Connolly H. C. Donaldson Hanna K. L. Emery J. P. Enos H. L. Fornasier S. Haberle C. W. Hanna R. D. Howell E. S. Kaplan H. H. Keller L. P. Lantz C. Li J. Y. Lim L. F. McCoy T. J. Merlin F. Nolan M. C. Praet A. Rozitis B. Sandford S. A. Schrader D. L. Thomas C. A. Zou X. D. Lauretta D. S. Highsmith D. E. Small J. Vokrouhlický D. Bowles N. E. Brown E. Donaldson Hanna K. L. Warren T. Brunet C. Chicoine R. A. Desjardins S. Gaudreau D. Haltigin T. Millington-Veloza S. Rubi A. Aponte J. Gorius N. Lunsford A. Allen B. Grindlay J. Guevel D. Hoak D. Hong J. Schrader D. L. Bayron J. Golubov O. Sánchez P. Stromberg J. Hirabayashi M. Hartzell C. M. Oliver S. Rascon M. Harch A. Joseph J. Squyres S. Richardson D. Emery J. P. McGraw L. Ghent R. Binzel R. P. Asad M. M. A. Johnson C. L. Philpott L. Susorney H. C. M. Cloutis E. A. Hanna R. D. Connolly H. C. Ciceri F. Hildebrand A. R. Ibrahim E. M. Breitenfeld L. Glotch T. Rogers A. D. Clark B. E. Ferrone S. Thomas C. A. Campins H. Fernandez Y. Chang W. Cheuvront A. Trang D. Tachibana S. Yurimoto H. Brucato J. R. Poggiali G. Pajola M. Dotto E. Mazzotta Epifani E., et al. 2019. Evidence for widespread hydrated minerals on asteroid (101955) Bennu. *Nature Astronomy* 3:332-340.

Hewins R. H., Bourot-Denise M., Zanda B., Leroux H., Barrat J.-A., Humayun M., Göpel C., Greenwood R. C., Franchi I. A., Pont S., Lorand J.-P., Cournède C., Gattacceca J., Rochette P., Kuga M., Marrocchi Y., and Marty B. 2014. The Paris meteorite, the least altered CM chondrite so far. *Geochimica et Cosmochimica Acta* 124:190-222.

Hoppe P., Leitner J., and Kodolányi J. 2015. New constraints on the abundances of silicate and oxide stardust from supernovae in the Acfer 094 meteorite. *The Astrophysical Journal Letters* 808.

Hoppe P., Leitner J., and Kodolányi J. 2017. The stardust abundance in the local interstellar cloud at the birth of the Solar System. *Nature Astronomy* 1:617-620.

Hoppe P., Strebel R., Eberhardt P., Amari S., and Lewis R. S. 2000. Isotopic properties of silicon carbide X grains from the Murchison meteorite in the size range 0.5-1.5 μm. *Meteoritics & Planetary Science* 35:1157–1176.

Howard K. T., Alexander C. M. O'D., Schrader D. L., and Dyl K. A. 2015. Classification of hydrous meteorites (CR, CM and C2 ungrouped) by phyllosilicate fraction: PSD-XRD modal mineralogy and planetesimal environments. *Geochimica et Cosmochimica Acta* 149:206-222.

Huss G. R. and Lewis R. S. 1995. Presolar diamond, SiC, and graphite in primitive chondrites: Abundances as a function of meteorite class and petrologic type. *Geochimica et Cosmochimica Acta* 59:115–160.

Huss G. R., Meshik A. P., Smith J. B., and Hohenberg C. M. 2003. Presolar diamond, silicon carbide, and graphite in carbonaceous chondrites: implications for thermal processing in the solar nebula. *Geochimica et Cosmochimica Acta* 67:4823–4848.

Hutcheon I. D., Huss G. R., Fahey A. J., and Wasserburg G. J. 1994. Extreme $^{26}$Mg and $^{17}$O enrichments in an Orgueil corundum: Identification of an interstellar oxide grain. *The Astrophysical Journal* 425:L97–L100.

Jacquet E., Barrat J.-A., Beck P., Caste F., Gattacceca J., Sonzogni C., and Gounelle M. 2016. Northwest Africa 5958: A weakly altered CM-related ungrouped chondrite, not a CI3. *Meteoritics and Planetary Science* 51:851-869.

Jones C., Fike D. A., and Peres P. 2017. Investigation of the quasi-simultaneous arrival (QSA) effect on a CAMECA IMS 7f-GEO. *Rapid Communications in Mass Spectrometry* 31:623-630.

Keller L. P. and Messenger S. 2011. On the origins of GEMS grains. *Geochimica et Cosmochimica Acta* 75:5336–5365.

Kimura M., Imae N., Komatsu M., Barrat J. A., Greenwood R. C., Yamaguchi A., and Noguchi T. 2020. The most primitive CM chondrites, Asuka 12085, 12169, and 12236, of subtypes 3.0–2.8: Their characteristic features and classification. *Polar Science* in press.

Kimura M., Imae N., Yamaguchi A., Greenwood R. C., Komatsu M., and Noguchi T. (2019) Primitive CM-related chondrites: Their characteristic features and classification. In *82nd Annual Meeting of The Meteoritical Society*, Abstract 6042.

Kitazato K., Milliken R. E., Iwata T., Abe M., Ohtake M., Matsuura S., Arai T., Nakauchi Y., Nakamura T., Matsuoka M., Senshu H., Hirata N., Hiroi T., Pilorget C., Brunetto R., Poulet F., Riu L., Bibring J.-P., Takir D., Domingue D. L., Vilas F., Barucci M. A., Perna D., Palomba E., Galiano A., Tsumura K., Osawa T., Komatsu M., Nakato A., Arai T., Takato N., Matsunaga T., Takagi Y., Matsumoto K., Kouyama T., Yokota Y., Tatsumi E., Sakatani N., Yamamoto Y., Okada T., Sugita S., Honda R., Morota T., Kameda S., Sawada H., Honda C., Yamada M., Suzuki H., Yoshioka K., Hayakawa M., Ogawa K., Cho Y., Shirai K., Shimaki Y., Hirata N., Yamaguchi A., Ogawa N., Terui F., Yamaguchi T., Takei Y., Saiki T., Nakazawa S., Tanaka S., Yoshikawa M., Watanabe S., and Tsuda Y. 2019. The surface composition of asteroid 162173 Ryugu from Hayabusa2 near-infrared spectroscopy. *Science* 364:272-275.

Le Guillou C. and Brearley A. 2014. Relationships between organics, water and early stages of aqueous alteration in the pristine CR3.0 chondrite MET 00426. *Geochimica et Cosmochimica Acta* 131:344–367.

Leitner J., Hoppe P., Floss C., Hillion F., and Henkel T. 2018. Correlated nanoscale characterization of a unique complex oxygen-rich stardust grain: Implications for circumstellar dust formation. *Geochimica et Cosmochimica Acta*. 221:255–274.

Leitner J., Metzler K., Vollmer C., Floss C., Haenecour P., Kodolányi J., Harries D., and Hoppe P. 2020. The presolar grain inventory of fine-grained chondrule rims in the Mighei-type (CM) chondrites. *Meteoritics & Planetary Science* 55:1176-1206.

Leitner J., Vollmer C., Hoppe P., and Zipfel J. 2012. Characterization of presolar material in the CR chondrite Northwest Africa 852. *The Astrophysical Journal* 745:38.

Leroux H., Cuvillier P., Zanda B., and Hewins R. H. 2015. GEMS-like material in the matrix of the Paris meteorite and the early stages of alteration of CM chondrites. *Geochimica et Cosmochimica Acta* 170:247-265.

Liu N., Gallino R., Cristallo S., Bisterzo S., Davis A. M., Trappitsch R., and Nittler L. R. 2018. New constraints on the major neutron source in low-mass AGB stars. *The Astrophysical Journal Letters* 865:112 (14pp).

Marrocchi Y., Gounelle M., Blanchard I., Caste F., and Kearsley A. T. 2014. The Paris CM chondrite: Secondary minerals and asteroidal processing. *Meteoritics & Planetary Science* 49:1232-1249.

Mostefaoui S. and Hoppe P. 2004. Discovery of abundant in situ silicate and spinel grains from red giant stars in a primitive meteorite. *The Astrophysical Journal* 613:L149–L152.

Mostefaoui S. 2011. The Search for Presolar Oxides in Paris. *Meteoritics and Planetary Science Supplement* 74.

Nguyen A. N., Nittler L. R., Stadermann F. J., Stroud R. M., and Alexander C. M. O'D. 2010. Coordinated analyses of presolar grains in the Allan Hills 77307 and Queen Elizabeth Range 99177 meteorites. *The Astrophysical Journal* 719:166–189.

Nguyen A. N., Stadermann F. J., Zinner E., Stroud R. M., Alexander C. M. O'D., and Nittler L. R. 2007. Characterization of presolar silicate and oxide grains in primitive carbonaceous chondrites. *The Astrophysical Journal* 656:1223–1240.

Nittler L. R., Alexander C. M. O'D., Davidson J., Riebe M. E. I., Stroud R. M., and Wang J. 2018. High abundances of presolar grains and $^{15}$N-rich organic matter in CO3.0 chondrite Dominion Range 08006. *Geochimica et Cosmochimica Acta* 226:107-131.

Nittler L. R., Alexander C. M. O'D., Foustoukos D., Patzer A., and Verdier-Paoletti M. J. 2020a. Asuka 12236, the most pristine CM chondrite to date. *Lunar and Planetary Science Conference* 51:Abstract #2276.

Nittler L. R., Alexander C. M. O'D., Gallino R., Hoppe P., Nguyen A., Stadermann F., and Zinner E. K. 2008. Aluminum-, calcium- and titanium-rich oxide stardust in ordinary chondrite meteorites. *The Astrophysical Journal* 682:1450–1478.

Nittler L. R., Alexander C. M. O'D., Gao X., Walker R. M., and Zinner E. 1997. Stellar sapphires: The properties and origins of presolar $Al_2O_3$ in meteorites. *The Astrophysical Journal* 483:475–495.

Nittler L. R. and Ciesla F. 2016. Astrophysics with extraterrestrial materials. *Annual Review of Astronomy and Astrophysics* 54:53–93.

Nittler L. R., Stroud R. M., Alexander C. M. O'D., and Howell K. 2019. Presolar grains in primitive ungrouped carbonaceous chondrite Northwest Africa 5958. *Meteoritics and Planetary Science* 55:1160-1175.

Nittler L. R., Verdier-Paoletti M. J., and Alexander C. M. O'D. (2020b) Microscale hydrogen and nitrogen isotopic distributions in pristine CM chondrite Asuka 12236. *Goldschmidt Conference*, https://doi.org/10.46427/gold2020.1938

Noguchi T., Yasutake M., Tsuchiyama A., Miyake A., Kimura M., Yamaguchi A., Imae N., Uesugi K., and Takeuchi A. 2020. Matrix mineralogy of the least altered CM-related chondrite Asuka 12169. *Lunar and Planetary Science Conference 51*:Abstract #1666.

Riebe M. E. I., Busemann H., Alexander C. M. O'D., Nittler L. R., Herd C. D. K., Maden C., Wang J., and Wieler R. 2019. Effects of aqueous alteration on primordial noble gases and presolar SiC in the carbonaceous chondrite Tagish Lake. *Meteoritics and Planetary Science* 55: 1257-1280

Rubin A. E., Trigo-Rodríguez J. M., Huber H., and Wasson J. T. 2007. Progressive aqueous alteration of CM carbonaceous chondrites. *Geochimica et Cosmochimica Acta* 71:2361-2382.

Slodzian G., Hillion F., Stadermann F. J., and Zinner E. 2004. QSA influences on isotopic ratio measurements. *Applied Surface Science* 231–232:874–877.

Tsuchiyama A., Noguchi T., Yasutake M., Miyake A., Kimura M., Yamaguchi A., Imae N., Uesugi K., and Takeuchi A. 2020. Three-dimensional nano/microtexture of a least altered CM-related chondrite Asuka 12169. *Lunar and Planetary Science Conference* 51:Abstract #1801.

Verdier-Paoletti M. J., Nittler L. R., and Wang J. 2019. First detection of presolar grains in Paris, The most preserved CM chondrite. *Lunar and Planetary Science Conference* 50:Abstract #2948.

Verdier-Paoletti M. J., Nittler L. R., and Wang J. 2020. New estimation of presolar grain abundances in the Paris meteorite. *Lunar and Planetary Science Conference* 51: Abstract #2523.

Vollmer C., Hoppe P., Stadermann F. J., Floss C., and Brenker F. E. 2009. NanoSIMS analysis and Auger electron spectroscopy of silicate and oxide stardust from the carbonaceous chondrite Acfer 094. *Geochimica et Cosmochimica Acta* 73:7127–7149.

Xu Y., Lin Y., Hao J., and Kimura M. (2020) The high abundances of presolar grains found in the most primitive CM meteorite, Asuka 12169. *11th Symposium on Polar Science*, National Institute of Polar Research, Japan, Abstract OAo5.

Zhao X., Floss C., Lin Y., and Bose M. 2013. Stardust investigation into the CR chondrite Grove Mountain 021710. *The Astrophysical Journal* 769:49 (16pp).

Zhao X., Lin Y., Yin Q.-Z., Zhang J., Hao J., Zolensky M., and Jenniskens P. 2014. Presolar grains in the CM2 chondrite Sutter's Mill. *Meteoritics and Planetary Science* 49:2038-2046.

Zinner E. 2014. 1.4 - Presolar grains. In *Meteorites and Cosmochemical Processes (Vol. 1), Treatise on Geochemistry (Second Edition, eds: H. D. Holland and K. K. Turekian)* (ed. A. M. Davis), pp. 181–213. Elsevier-Pergamon, Oxford.

Zinner E., Nittler L. R., Gallino R., Karakas A. I., Lugaro M., Straniero O., and Lattanzio J. C. 2006. Silicon and carbon isotopic ratios in AGB stars: SiC grain data, models, and the Galactic evolution of the Si isotopes. *The Astrophysical Journal* 650:350–373.

Table 1. Analyzed areas and presolar grain abundance results for A12236 and A12169 meteorites.

| Meteorite | Region | Total area ($\mu m^2$) | O-rich presolar grains # | O-rich presolar grains Abundance (ppm) | Presolar SiC grains # | Presolar SiC grains Abundance (ppm) |
|---|---|---|---|---|---|---|
| A12236 | | | | | | |
| | R1 | 2500 | 2 | $40^{+53}_{-26}$ | 0 | 0 |
| | R6 | 3300 | 5 | $80^{+54}_{-34}$ | 1 | $21^{+48}_{-17}$ |
| | R7 | 2490 | 2 | $30^{+39}_{-19}$ | 1 | $37^{+85}_{-30}$ |
| | R10 | 3200 | 2 | $25^{+33}_{-16}$ | 2 | $21^{+41}_{-20}$ |
| | R11 | 1940 | 2 | $50^{+66}_{-33}$ | 1 | $28^{+64}_{-23}$ |
| | R13 | 2800 | 5 | $93^{+63}_{-40}$ | 1 | $6^{+14}_{-5}$ |
| | Total | 16,230 | 18 | $58^{+18}_{-12}$ | 6 | $20^{+12}_{-8}$ |
| A12169 | | | | | | |
| | R1 | 6470 | 24 | $371^{+92}_{-75}$ | 6 | $46^{+27}_{-18}$ |
| | R2 | 11100 | 40 | $219^{+40}_{-34}$ | 13 | $61^{+22}_{-17}$ |
| | R3 | 3330 | 15 | $251^{+83}_{-64}$ | 2 | $28^{+37}_{-18}$ |
| | R4 | 2610 | 11 | $308^{+124}_{-91}$ | 4 | $66^{+52}_{-32}$ |
| | total | 23,510 | 90 | $275^{+55}_{-50}{}^{+120}_{-90}$ | 25 | $62^{+15}_{-12}$ |
| | total (wo big) | | | $236^{+37}_{-34}{}^{+79}_{-65}$ | | |

All errors are 1-σ except that 1- and 2-σ errors are shown for the total O-rich presolar grain abundance in A12169.

Table 2. Presolar O-rich grains identified in A12236 and A12169 meteorites (1-σ errors).

| Grain* | Region | Size (nm) | Group | $^{17}O/^{16}O$ (×10$^{-4}$) | $^{18}O/^{16}O$ (×10$^{-3}$) | Si$^-$/O$^-$ | AlO$^-$/O$^-$ | Phase |
|---|---|---|---|---|---|---|---|---|
| A12236-1 | R1 | 238 | 1 | 12.61 ± 1.06 | 1.785 ± 0.129 | 0.015 | 0.005 | Silicate |
| A12236-2 | R1 | 303 | 1 | 8.92 ± 0.72 | 2.142 ± 0.108 | 0.018 | 0.002 | Silicate |
| A12236-3 | R6 | 273 | 1 | 7.73 ± 0.62 | 1.845 ± 0.096 | 0.012 | 0.005 | Silicate |
| A12236-4 | R6 | 219 | 1 | 6.81 ± 0.71 | 2.328 ± 0.127 | 0.013 | 0.005 | Silicate |
| A12236-5 | R6 | 324 | 1 | 5.80 ± 0.47 | 1.380 ± 0.086 | 0.018 | 0.004 | Silicate |
| A12236-6 | R6 | 296 | 4 | 4.33 ± 0.44 | 2.887 ± 0.109 | 0.013 | 0.018 | Silicate |
| A12236-7 | R6 | 264 | 1 | 6.42 ± 0.57 | 2.010 ± 0.099 | 0.014 | 0.005 | Silicate |
| A12236-8 | R7 | 229 | 1 | 8.67 ± 1.05 | 1.848 ± 0.128 | 0.017 | 0.003 | Silicate |
| A12236-9 | R7 | 247 | 1 | 6.16 ± 0.63 | 1.384 ± 0.130 | 0.017 | 0.003 | Silicate |
| A12236-10 | R10 | 256 | 1 | 9.92 ± 0.92 | 2.073 ± 0.129 | 0.020 | 0.004 | Silicate |
| A12236-11 | R10 | 238 | 3 | 3.47 ± 0.62 | 1.300 ± 0.139 | 0.018 | 0.004 | Silicate |
| A12236-12 | R11 | 317 | 1 | 12.06 ± 0.90 | 1.905 ± 0.111 | 0.024 | 0.003 | Silicate |

| Grain | Region | Size (nm) | n | ¹²C/¹³C | ¹⁴N/¹⁵N | ²⁶Al/²⁷Al | ³⁰Si/²⁸Si | Type |
|---|---|---|---|---|---|---|---|---|
| A12236-13 | R11 | 209 | 4 | 3.91 ± 0.75 | 2.956 ± 0.198 | 0.018 | 0.003 | Silicate |
| A12236-14 | R13 | 296 | 1 | 8.54 ± 0.67 | 1.793 ± 0.100 | 0.020 | 0.001 | Silicate |
| A12236-15 | R13 | 198 | 1 | 3.86 ± 0.64 | 1.248 ± 0.141 | 0.016 | 0.006 | Silicate |
| A12236-16 | R13 | 380 | 1 | 16.44 ± 0.77 | 1.984 ± 0.083 | 0.022 | 0.001 | Silicate |
| A12236-17 | R13 | 247 | 1 | 6.51 ± 0.74 | 1.663 ± 0.126 | 0.013 | 0.006 | Silicate |
| A12236-18 | R13 | 219 | 4 | 4.48 ± 0.67 | 3.001 ± 0.169 | 0.020 | 0.006 | Silicate |
| | | | | | | | | |
| A12169-1 | R1 | 224 | 1 | 6.90 ± 0.52 | 1.933 ± 0.084 | 0.020 | 0.003 | Silicate |
| A12169-2 | R1 | 181 | 1 | 6.34 ± 0.48 | 2.016 ± 0.082 | 0.015 | 0.008 | Silicate |
| A12169-3 | R1 | 652 | 1 | 5.39 ± 0.17 | 1.900 ± 0.032 | 0.017 | 0.007 | Silicate |
| A12169-4 | R1 | 211 | 1 | 6.43 ± 0.42 | 1.580 ± 0.070 | 0.015 | 0.005 | Silicate |
| A12169-5 | R1 | 201 | 1 | 6.51 ± 0.48 | 2.083 ± 0.082 | 0.014 | 0.007 | Silicate |
| A12169-6 | R1 | 256 | 1 | 6.88 ± 0.39 | 1.912 ± 0.065 | 0.015 | 0.038 | Silicate |
| A12169-7 | R1 | 164 | 1 | 5.22 ± 0.50 | 1.512 ± 0.094 | 0.014 | 0.014 | Silicate |
| A12169-8 | R1 | 399 | 1 | 8.69 ± 0.33 | 1.877 ± 0.048 | 0.021 | 0.002 | Silicate |
| A12169-9 | R1 | 206 | 1 | 7.16 ± 0.51 | 1.945 ± 0.082 | 0.017 | 0.006 | Silicate |
| A12169-10 | R1 | 472 | 4 | 7.53 ± 0.35 | 4.372 ± 0.080 | 0.0021 | 0.037 | Oxide |
| A12169-11 | R1 | 158 | 1 | 6.70 ± 0.55 | 1.784 ± 0.091 | 0.018 | 0.022 | Silicate |
| A12169-12 | R1 | 158 | 1 | 5.48 ± 0.53 | 1.483 ± 0.096 | 0.017 | 0.004 | Silicate |
| A12169-13 | R1 | 241 | 1 | 5.83 ± 0.41 | 1.904 ± 0.072 | 0.025 | 0.002 | Silicate |
| A12169-14 | R1 | 224 | 1 | 6.92 ± 0.45 | 2.001 ± 0.074 | 0.016 | 0.012 | Silicate |
| A12169-15 | R1 | 268 | 1 | 8.03 ± 0.46 | 1.810 ± 0.069 | 0.019 | 0.003 | Silicate |
| A12169-16 | R1 | 404 | 1 | 5.09 ± 0.28 | 1.964 ± 0.054 | 0.020 | 0.003 | Silicate |
| A12169-17 | R1 | 355 | 1 | 11.38 ± 0.41 | 1.807 ± 0.052 | 0.016 | 0.011 | Silicate |
| A12169-18 | R1 | 264 | 1 | 8.67 ± 0.44 | 1.930 ± 0.065 | 0.015 | 0.007 | Silicate |
| A12169-19 | R1 | 249 | 1 | 12.57 ± 0.55 | 1.927 ± 0.067 | 0.018 | 0.007 | Silicate |
| A12169-20 | R1 | 308 | 1 | 5.81 ± 0.33 | 1.851 ± 0.058 | 0.023 | 0.005 | Silicate |
| A12169-21 | R1 | 233 | 1 | 11.55 ± 0.58 | 1.956 ± 0.074 | 0.020 | 0.005 | Silicate |
| A12169-22 | R1 | 175 | 4 | 4.95 ± 0.51 | 2.957 ± 0.122 | 0.018 | 0.004 | Silicate |
| A12169-23 | R1 | 1092 | 1 | 31.48 ± 0.28 | 1.896 ± 0.020 | 0.0064 | 0.036 | Hibonite |
| A12169-24 | R1 | 152 | 1 | 5.62 ± 0.54 | 1.247 ± 0.096 | 0.022 | 0.008 | Silicate |
| A12169-25 | R2 | 233 | 4 | 3.86 ± 0.31 | 3.074 ± 0.085 | 0.012 | 0.009 | Silicate |
| A12169-26 | R2 | 211 | 1 | 5.62 ± 0.46 | 1.778 ± 0.085 | 0.019 | 0.005 | Silicate |
| A12169-27 | R2 | 371 | 1 | 5.36 ± 0.25 | 1.936 ± 0.046 | 0.015 | 0.008 | Silicate |
| A12169-28 | R2 | 201 | 1 | 5.51 ± 0.47 | 1.323 ± 0.085 | 0.021 | 0.003 | Silicate |
| A12169-29 | R2 | 201 | 1 | 6.15 ± 0.55 | 1.389 ± 0.095 | 0.017 | 0.002 | Silicate |
| A12169-30 | R2 | 245 | 1 | 7.18 ± 0.45 | 1.966 ± 0.072 | 0.015 | 0.007 | Silicate |
| A12169-31 | R2 | 256 | 1 | 6.24 ± 0.40 | 1.902 ± 0.069 | 0.019 | 0.003 | Silicate |
| A12169-32 | R2 | 233 | 1 | 6.42 ± 0.45 | 1.958 ± 0.076 | 0.018 | 0.005 | Silicate |
| A12169-33 | R2 | 314 | 1 | 8.60 ± 0.42 | 2.220 ± 0.065 | 0.017 | 0.010 | Oxide |
| A12169-34 | R2 | 292 | 1 | 6.52 ± 0.39 | 1.136 ± 0.065 | 0.024 | 0.004 | Silicate |
| A12169-35 | R2 | 201 | 1 | 7.56 ± 0.58 | 1.362 ± 0.095 | 0.018 | 0.004 | Silicate |
| A12169-36 | R2 | 326 | 1 | 6.65 ± 0.32 | 1.880 ± 0.053 | 0.015 | 0.011 | Oxide |
| A12169-37 | R2 | 158 | 4 | 3.56 ± 0.44 | 2.540 ± 0.113 | 0.016 | 0.008 | Silicate |
| A12169-38 | R2 | 275 | 1 | 6.78 ± 0.41 | 1.535 ± 0.067 | 0.016 | 0.004 | Silicate |
| A12169-39 | R2 | 260 | 4 | 5.29 ± 0.38 | 3.039 ± 0.088 | 0.022 | 0.005 | Silicate |
| A12169-40 | R2 | 741 | 1 | 9.57 ± 0.22 | 1.916 ± 0.030 | 0.023 | 0.001 | Silicate |
| A12169-41 | R2 | 329 | 1 | 5.47 ± 0.30 | 1.877 ± 0.054 | 0.021 | 0.003 | Silicate |
| A12169-42 | R2 | 206 | 1 | 6.65 ± 0.48 | 1.867 ± 0.080 | 0.018 | 0.006 | Silicate |
| A12169-43 | R2 | 508 | 2 | 7.56 ± 0.26 | 0.755 ± 0.040 | 0.028 | 0.001 | Silicate |
| A12169-44 | R2 | 164 | 1 | 6.03 ± 0.69 | 1.351 ± 0.120 | 0.019 | 0.002 | Silicate |
| A12169-45 | R2 | 237 | 1 | 5.64 ± 0.35 | 1.449 ± 0.071 | 0.019 | 0.010 | Silicate |
| A12169-46 | R2 | 145 | 1 | 4.20 ± 0.58 | 1.368 ± 0.120 | 0.025 | 0.003 | Silicate |
| A12169-47 | R2 | 271 | 1 | 3.76 ± 0.47 | 1.404 ± 0.106 | 0.025 | 0.000 | Silicate |
| A12169-48 | R2 | 289 | 1 | 7.80 ± 0.40 | 1.892 ± 0.061 | 0.019 | 0.006 | Silicate |
| A12169-49 | R2 | 191 | 1 | 6.78 ± 0.55 | 1.823 ± 0.091 | 0.017 | 0.006 | Silicate |
| A12169-50 | R2 | 249 | 1 | 5.57 ± 0.38 | 1.900 ± 0.069 | 0.021 | 0.007 | Silicate |
| A12169-51 | R2 | 338 | 1 | 15.32 ± 0.54 | 1.883 ± 0.060 | 0.026 | 0.002 | Silicate |
| A12169-52 | R2 | 260 | 1 | 6.06 ± 0.40 | 1.941 ± 0.069 | 0.024 | 0.008 | Silicate |
| A12169-53 | R2 | 215 | 1 | 5.92 ± 0.44 | 1.934 ± 0.079 | 0.016 | 0.009 | Silicate |
| A12169-54 | R2 | 145 | 1 | 7.46 ± 0.71 | 2.039 ± 0.113 | 0.020 | 0.008 | Silicate |
| A12169-55 | R2 | 256 | 1 | 4.72 ± 0.35 | 1.309 ± 0.070 | 0.030 | 0.005 | Silicate |
| A12169-56 | R2 | 241 | 1 | 10.54 ± 0.62 | 1.741 ± 0.083 | 0.0091 | 0.042 | Oxide |
| A12169-57 | R2 | 335 | 1 | 3.69 ± 0.24 | 1.417 ± 0.054 | 0.011 | 0.063 | Silicate |
| A12169-58 | R2 | 181 | 4 | 3.85 ± 0.39 | 3.042 ± 0.108 | 0.021 | 0.007 | Silicate |
| A12169-59 | R2 | 228 | 1 | 7.83 ± 0.41 | 2.065 ± 0.065 | 0.005 | 0.005 | Silicate |
| A12169-60 | R2 | 186 | 1 | 7.06 ± 0.55 | 2.084 ± 0.092 | 0.025 | 0.005 | Silicate |
| A12169-61 | R2 | 206 | 4 | 4.81 ± 0.43 | 2.960 ± 0.103 | 0.019 | 0.005 | Silicate |
| A12169-62 | R2 | 186 | 1 | 6.53 ± 0.58 | 1.984 ± 0.099 | 0.023 | 0.036 | Silicate |
| A12169-63 | R2 | 298 | 1 | 5.32 ± 0.32 | 1.556 ± 0.060 | 0.021 | 0.019 | Silicate |
| A12169-64 | R2 | 74 | 4 | 3.31 ± 0.66 | 2.788 ± 0.174 | 0.016 | 0.009 | Silicate |
| A12169-65 | R3 | 228 | 1 | 7.26 ± 0.55 | 1.999 ± 0.089 | 0.017 | 0.004 | Silicate |
| A12169-66 | R3 | 228 | 1 | 7.23 ± 0.51 | 1.770 ± 0.081 | 0.014 | 0.007 | Silicate |
| A12169-67 | R3 | 360 | 1 | 6.48 ± 0.37 | 1.940 ± 0.063 | 0.017 | 0.003 | Silicate |
| A12169-68 | R3 | 320 | 1 | 6.29 ± 0.36 | 2.142 ± 0.064 | 0.016 | 0.006 | Silicate |
| A12169-69 | R3 | 107 | 1 | 5.40 ± 0.74 | 2.660 ± 0.159 | 0.016 | 0.012 | Silicate |
| A12169-70 | R3 | 302 | 2 | 9.47 ± 0.45 | 0.831 ± 0.064 | 0.023 | 0.009 | Silicate |
| A12169-71 | R3 | 332 | 1 | 12.13 ± 1.04 | 1.747 ± 0.063 | 0.018 | 0.004 | Silicate |
| A12169-72 | R3 | 278 | 1 | 8.78 ± 0.47 | 2.058 ± 0.070 | 0.017 | 0.006 | Silicate |
| A12169-73 | R3 | 374 | 1 | 3.68 ± 0.28 | 1.653 ± 0.062 | 0.027 | 0.001 | Silicate |
| A12169-74 | R3 | 341 | 1 | 9.18 ± 0.42 | 1.813 ± 0.060 | 0.019 | 0.005 | Silicate |
| A12169-75 | R3 | 224 | 4 | 3.23 ± 0.37 | 2.484 ± 0.093 | 0.020 | 0.004 | Silicate |
| A12169-76 | R3 | 97 | 1 | 7.70 ± 0.89 | 1.698 ± 0.141 | 0.017 | 0.009 | Silicate |
| A12169-77 | R3 | 237 | 1 | 7.83 ± 0.60 | 2.142 ± 0.097 | 0.015 | 0.006 | Silicate |
| A12169-78 | R3 | 191 | 1 | 7.39 ± 0.63 | 1.981 ± 0.099 | 0.019 | 0.005 | Silicate |
| A12169-79 | R3 | 170 | 1 | 7.10 ± 0.72 | 1.716 ± 0.116 | 0.025 | 0.010 | Silicate |

| Grain | Region | Size (nm) | | $\delta^{13}C$ (‰) | $^{12}C/^{13}C$ | | | |
|---|---|---|---|---|---|---|---|---|
| A12169-80 | R4 | 170 | 4 | 3.76 ± 0.43 | 2.540 ± 0.107 | 0.014 | 0.005 | Silicate |
| A12169-81 | R4 | 249 | 4 | 4.23 ± 0.39 | 2.961 ± 0.099 | 0.020 | 0.008 | Silicate |
| A12169-82 | R4 | 201 | 1 | 6.49 ± 0.48 | 1.875 ± 0.081 | 0.016 | 0.006 | Silicate |
| A12169-83 | R4 | 164 | 1 | 7.34 ± 0.74 | 1.963 ± 0.119 | 0.021 | 0.001 | Silicate |
| A12169-84 | R4 | 502 | 2 | 6.76 ± 0.23 | 0.190 ± 0.095 | 0.025 | 0.003 | Silicate |
| A12169-85 | R4 | 115 | 1 | 7.07 ± 0.70 | 1.951 ± 0.114 | 0.017 | 0.009 | Silicate |
| A12169-86 | R4 | 314 | 5 | 80.75 ± 3.15 | 1.802 ± 0.054 | 0.020 | 0.003 | Silicate |
| A12169-87 | R4 | 338 | 1 | 5.35 ± 0.26 | 1.936 ± 0.049 | 0.016 | 0.006 | Silicate |
| A12169-88 | R4 | 191 | 4 | 4.71 ± 0.46 | 2.613 ± 0.105 | 0.015 | 0.020 | Silicate |
| A12169-89 | R4 | 233 | 1 | 6.82 ± 0.44 | 1.911 ± 0.073 | 0.014 | 0.016 | Silicate |
| A12169-90 | R4 | 573 | 1 | 4.42 ± 0.14 | 1.883 ± 0.029 | 0.013 | 0.010 | Silicate |

*Grain name indicates in which meteorite a grain was identified

Table 3. Presolar SiC grains identified in A12236 and A12169 meteorites (1-σ errors)

| Grain* | Region | Size (nm) | $\delta^{13}C$ (‰) | $^{12}C/^{13}C$ | Si-/C- |
|---|---|---|---|---|---|
| A12236-SiC-1 | R6 | 300 | 915 ± 170 | 46.5 ± 4.5 | 1.55 |
| A12236-SiC-2 | R7 | 340 | 715 ± 76 | 51.9 ± 2.4 | 0.21 |
| A12236-SiC-3 | R10 | 254 | 3254 ± 295 | 20.9 ± 1.6 | 0.74 |
| A12236-SiC-4 | R10 | 206 | 595 ± 117 | 55.8 ± 4.4 | 0.23 |
| A12236-SiC-5 | R11 | 245 | 995 ± 167 | 44.6 ± 4.1 | 0.38 |
| A12236-SiC-6 | R13 | 144 | 407 ± 95 | 63.3 ± 4.6 | 0.02 |
| | | | | | |
| A12169-SiC-1 | R1 | 164 | 481 ± 072 | 60.1 ± 3.1 | 0.15 |
| A12169-SiC-2 | R1 | 253 | 1107 ± 155 | 42.2 ± 3.4 | 0.92 |
| A12169-SiC-3 | R1 | 220 | 3900 ± 267 | 18.2 ± 1.0 | 1.00 |
| A12169-SiC-4 | R1 | 335 | 969 ± 146 | 45.2 ± 3.6 | 0.99 |
| A12169-SiC-5 | R1 | 237 | 648 ± 082 | 54.0 ± 2.8 | 0.33 |
| A12169-SiC-6 | R1 | 268 | 1635 ± 114 | 33.8 ± 1.5 | 0.44 |
| A12169-SiC-7 | R2 | 347 | 961 ± 044 | 45.4 ± 1.0 | 0.98 |
| A12169-SiC-8 | R2 | 268 | 1268 ± 096 | 39.2 ± 1.7 | 0.94 |
| A12169-SiC-9 | R2 | 268 | 1268 ± 096 | 39.2 ± 1.7 | 0.94 |
| A12169-SiC-10 | R2 | 289 | 0623 ± 055 | 54.8 ± 1.9 | 0.81 |
| A12169-SiC-11 | R2 | 220 | 2316 ± 223 | 26.8 ± 1.9 | 0.94 |
| A12169-SiC-12 | R2 | 289 | 941 ± 101 | 45.8 ± 2.5 | 0.78 |
| A12169-SiC-13 | R2 | 245 | 426 ± 053 | 62.4 ± 2.4 | 0.71 |
| A12169-SiC-14 | R2 | 228 | 905 ± 139 | 46.7 ± 3.7 | 0.99 |
| A12169-SiC-15 | R2 | 215 | 639 ± 115 | 54.3 ± 4.1 | 1.10 |
| A12169-SiC-16 | R2 | 256 | 411 ± 076 | 63.1 ± 3.6 | 1.08 |
| A12169-SiC-17 | R2 | 206 | 497 ± 099 | 59.4 ± 4.2 | 0.35 |
| A12169-SiC-18 | R2 | 206 | 328 ± 081 | 67.0 ± 4.3 | 1.09 |
| A12169-SiC-19 | R2 | 298 | -390 ± 067 | 145.9 ± 18.1 | 0.43 |

| Grain | | | | | |
|---|---|---|---|---|---|
| A12169-SiC-20 | R3 | 224 | $-267 \pm 058$ | $121.4 \pm 10.5$ | 1.20 |
| A12169-SiC-21 | R3 | 264 | $-282 \pm 049$ | $123.9 \pm 9.1$ | 0.26 |
| A12169-SiC-22 | R4 | 233 | $693 \pm 130$ | $52.6 \pm 4.4$ | 0.98 |
| A12169-SiC-23 | R4 | 241 | $3284 \pm 171$ | $20.8 \pm 0.9$ | 0.76 |
| A12169-SiC-24 | R4 | 191 | $1115 \pm 167$ | $42.1 \pm 3.6$ | 0.69 |
| A12169-SiC-25 | R4 | 268 | $4459 \pm 235$ | $16.3 \pm 0.7$ | 1.21 |

*Grain name indicates in which meteorite a grain was identified